\documentclass[12pt]{article}
\usepackage[cp1251]{inputenc}
\usepackage{epic}
\usepackage{eepic}
\usepackage{graphics}
\usepackage{amsfonts}
  \usepackage{amsthm}
 \usepackage{amsmath}
 \usepackage{amscd}
 \usepackage{amssymb}
 \usepackage{latexsym}
\usepackage{graphicx}
\usepackage{enumerate} 

 \numberwithin{equation}{section}
 \newtheorem{thm}{Theorem}[section]
 \newtheorem{lm}{Lemma}[section]
 \newtheorem{prop}{Proposition}[section]
 \theoremstyle{definition}
 \newtheorem{Cor}{Corollary}[section]
 
\newcommand{\ds}{\displaystyle}

\title{Globally Existing Solutions to the Problem of Dirichlet
for the Fractional $3D$ Poisson Equation}
\author{Toshko Boev  and   Georgi Georgiev \\
 Faculty of Mathematics and Informatics, \\
 Sofia University ``St. Kl. Ohridski'', \\
 5 James Bourchier Blvd., 1164 Sofia, Bulgaria}
\date{}

\begin{document}

\maketitle

\begin{abstract}
A common approach is present concerning the problem of Dirichlet, both for bounded $3D$ domains $\Omega$ and their (unbounded) complements $C \Omega=\mathbb{R}^3\setminus(\Omega \cup \partial\Omega)$, regarding the fractional ($3D$) Poisson equation $(-\Delta)^{\alpha/2}u=f$, with $\alpha$  assumed as $1<\alpha\le 2$.The solutions   $u(x)$ are sought from the Lauren Schwartz class   $S'=S'(\mathbb{R}^3)$-- of tempered distributions, so that -- applying on $u(x)$  the globally defined in $\mathbb{R}^3$  fractional Laplacian  $(-\Delta)^{\alpha/2}$   (by the Fourier transformation)-- to get valid the identity $(-\Delta)^{\alpha/2}u(x)\equiv f(x)$, respectively in $\Omega $ or $C \Omega$. The boundary condition $u|_{\Gamma}=\phi$, with $\Gamma=\partial\Omega$ -- the $2D$ contour of $\Omega $ is satisfied by the trace representation $B_{\alpha,\Gamma}[g]:=v_{\alpha,g}|_{\Gamma}$ of  the single layer Riesz type potential $v_{\alpha ,g}(x):=\int_{\Gamma}c_{3,\alpha}g(y)|x-y|^{\alpha-3}ds_y$, $x\in\mathbb{R}^3$, $g\in L_2(\Gamma)$, ($c_{3,\alpha}$ -- a corresponding constant). The key point here is the positive answer about whether $g=0$  is the unique solution of the equation $B_{\alpha,\Gamma}[g]=0$, where $B_{\alpha,\Gamma}\, :\, L_2(\Gamma)\rightarrow L_2(\Gamma)$ is a bounded (linear and compact) integral operator. The found solutions are expressed by the formula $u=v_{\alpha,g}+u_{\alpha, f}$, with $g$ depending on the boundary data $\phi$ and $u_{\alpha, f}$ -- the volume  Riesz potential, $u_{\alpha ,f}(x):=\int_{{\bf D}^0}c_{3,\alpha}f(y)|x-y|^{\alpha-3}dy$,  $x\in\mathbb{R}^3$, ${\bf D}^0=\Omega$ or ${\bf D}^0=C\Omega$. Several basic properties of solutions   are proved by this formula, in particular -- for their interior regularity and asymptotic behaviour at $|x|\rightarrow\infty$. Especial emphasizing deserves the property for the infinite regularity in $\Omega$  and  $C \Omega$ of the fractional harmonic functions $u(x)\, :\, (-\Delta)^{\alpha/2}u(x)\equiv0 $ (it becomes evident from the expression $u=v_{\alpha,g}+u_{\alpha,f}$).
\end{abstract}
\bigskip

\section{Introduction}
The topic for the fractional Laplacian has recently engaged an increasing interest (e.g. \cite{Lishke} and the big number of the relevant results cited there in). Nevertheless, the treatment for instance of the Dirichlet problem in the exterior of bounded, say $3D$, domains seems to be not analysed for the fractional Laplace (and Poisson) equation. And similar remarks can be made concerning the (infinite) interior regularity of the fractional harmonic functions (satisfying the equation  $(-\Delta)^{\alpha/2}u=0$), for a given domain $\mathcal{D}\subset\mathbb{R}^3$. Here we deal with $3D$ bounded domains $\Omega\subset\mathbb{R}^3$  and their complements $C \Omega=\mathbb{R}^3\setminus(\Omega \cup \Gamma)$, with $\Gamma=\partial\Omega$ -- the closed ($2D$) boundary surface of  $\Omega$, assumed of second order ( $C^2$) regularity. For the Laplacian degree $\alpha/2$ in 
$ (-\Delta)^{\alpha/2}$ we suppose here $1<\alpha\le2$ and the action $ (-\Delta)^{\alpha/2}u$ is defined by its Fourier transform $|\xi|^{\alpha}\hat{u}(\xi)$, for $u\in S'=S'(\mathbb{R}^3)$ -- the class of the Schwartz tempered distributions (e.g. \cite{Hermander}); $|\xi|$ is the length of the vector $\xi \in \mathbb{R}^3$ and $\hat{u}(\xi)=\mathbb{F}[u](\xi)$ is the Fourier image of  $u$. Concerning the Fourier transformation we proceed at the convention $\hat{\phi}(\xi)=\int_{\mathbb{R}^3}\exp(-i\langle x,\xi\rangle\phi (x)dx$ and 
$\phi(x)=\frac{1}{(2\pi)^3}\int_{\mathbb{R}^3}\exp(i\langle x,\xi\rangle\hat{\phi} (x)dx$, $\phi\in S=S(\mathbb{R}^3)$ -- the Schwartz class of the fast decreasing  (infinitely smooth) functions (\cite{Hermander}),  $\langle x,\xi\rangle $ is the scalar product of  the vectors $x,\,\xi$. Thus we have $(-\Delta)^{\alpha/2}u(x):=\mathbb{F}^{-1}[|\xi|^{\alpha}\hat{u}(\xi)](x)$, $u\in S'$, $x$   varying in $\mathbb{R}^3$, where $\mathbb{F}^{-1}$ is the inverse map of $\mathbb{F}$.  (The used here symbol  $(-\Delta)^{\alpha/2}$ for the fractional Laplacian follows the introduced one in \cite{Lishke}.)

Before to sketch the essence of the approach used, let us note via the possible applications that it admits also to consider a relevant treatment of the fractional Helmholtz equation related for instance to electrostatics of heterogeneous material systems. Our approach to the problem of Dirichlet is based on exploring the simple but effective idea to act with the global Laplacian, i.e $(-\Delta)^{\alpha/2}u=\mathbb{F}^{-1}[|.|^{\alpha}\mathbb{F}]$, on such distributions $u$  from $S'$   that the action product $(-\Delta)^{\alpha/2}u$ coincides on $\Omega$ with a prescribed function (distribution) $f$, and possess traces ($u|_{\Gamma}$) on $\Gamma$ (with prescribed values $\varphi(x)$  of $u|_{\Gamma}$, $x\in \Gamma$). Additionally noted, the way of looking for globally defined solutions (given a boundary value problem) has been firstly suggested from the $1D$ case of the Poisson equation  $(-\Delta)^{\alpha/2}u=f$, with $\Delta=\ds{\frac{d^2}{dx^2}}$, considered in unbounded intervals $l_0<x<\infty$, $l_0\in \mathbb{R}^1$. (We give in the Appendix some details to this case.)   

Slightly formalized, the above idea reads the following. Given a function $f(x)$, $x\in\Omega$, say bounded, i.e. $f\in L^{\infty}(\Omega)$, and boundary data $\varphi(x)$, $x\in \Gamma$ for instance $\varphi\in L_2(\Gamma)$,  the point is to solve the (extended on $\mathbb{R}^3$ ) equation $(-\Delta)^{\alpha/2}u=F^0$, with $F^0\in S' \, : \, F^0|_{\Omega}=f$, by a suitable distribution $u\in S' $, satisfying the condition $u|_{\Gamma}=\varphi $. (As seen below, the proper choice for $F^0$  is: $F^0|_{C\Omega }=0$.) If assume found such $u\in S'$, we could get a globally existing (i.e. defined on $\mathbb{R}^3$) solution of the problem under consideration, reformulated now in the form:

\begin{equation}
\label{0.1}
a)\, (-\Delta)^{\alpha/2} u|_{\Omega}=f,\, (u\in S'); \, b)\, u|_{\Omega}=\varphi .
\end{equation}
The above formulation actually gives the shortest illustration to the used here approach concerning the problem of Dirichlet (posed to given domain $\Omega $, in particular -- bounded), for the fractional Poisson equation.
We are close now to the key question for existence of solutions in $S'$, of equation (\ref{0.1}.a), at a sufficiently large class of boundary data. As a first accessory step to that goal, consider the volume type potential $U_{\beta ,f}:=\ds{\int_{\Omega}\frac{f(y)dy}{|x-y|^{\beta}}}$, $x\in\mathbb{R}^3$, $0<\beta$, $\beta=\beta(\alpha)$, prompted from the analogy with the conventional case of (\ref{0.1}.a), where $\alpha=2$, $\beta=1$. Following \cite{Lishke}, we will call such potentials Riesz (volume) potentials, and -- enlarging the terminology -- the functions of the type    $V_{\beta ,g}:=\ds{\int_{\Gamma}\frac{g(y)ds_y}{|x-y|^{\beta}}}$, ($x\in\mathbb{R}^3$) will be called surface (or single layer) Riesz potentials (here $g\in L_2(\Gamma)$, by assumption, and $ds_y$   is the known surface differential element). The right value of $\beta$  , namely $\beta=3-\alpha$, is however directly shown from the well known (e.g. \cite{Stein}, \cite{Landkof}, and also \cite{Lishke}) Fourier transform relation 
$\mathbb{F}\, :\, \ds{\frac{c_{3,\alpha}}{|x|^{3-\alpha}}}\rightarrow \ds{\frac{1}{|\xi |^{\alpha}}} $, with $c_{3,\alpha}=\ds{\frac{\Gamma(\frac{3-\alpha}{2})}{2^{\alpha}\pi^{3/2}\Gamma(\frac{\alpha}{2})}}$. And taking the precise expression instead of $U_{\beta,f}(x)$  we get the potential $u_{\alpha ,f}:=\ds{\int_{\Omega}\frac{c_{3,\alpha}f(y)dy}{|x-y|^{3-\alpha}}}$,  ($x\in\mathbb{R}^3$), which satisfies the equation (\ref{0.1}.a). Certainly, $u_{\alpha ,f}:=\ds{\int_{\mathbb{R}^3}\frac{c_{3,\alpha}f^0[f](y)dy}{|x-y|^{3-\alpha}}}=\ds{\frac{c_{3,\alpha}}{|.|^{3-\alpha}}}*f^0[f]$, where $f^0[f](y)=f(y)$ for $y\in\Omega$, and $f^0[f](y)=0$, when $y\in C\Omega$, $U*V$ is the convolution (see \cite{Hermander} for details) of $U\,, V\in S'$. Then $ (-\Delta)^{\alpha/2} u_{\alpha,f}=\left( (-\Delta)^{\alpha/2}\ds{\frac{c_{3,\alpha}}{|.|^{3-\alpha}}}\right)*f^0[f]=\delta * f^0=f^0$, here $\delta=\delta(x)$, ($x\in \mathbb{R}^3$) is the supported in the point $x=0$  Dirac delta function; i.e. $ \left((-\Delta)^{\alpha/2} u_{\alpha,f}\right)(x)=f^0[f](x)$, $x\in \mathbb{R}^3$ (therefore $(-\Delta)^{\alpha/2} u_{\alpha,f}|_{\Omega}=f$). These calculations evidently remain valid also for the generalized expression of $u_{\alpha,f}$,
\begin{equation}
\label{1.1}
u_{\alpha ,f}:=\int_{\mathcal{D}}\frac{c_{3,\alpha}f(y)dy}{|x-y|^{3-\alpha}},\, x\in \mathbb{R}^3, \, \mathcal{D}=\Omega\,\, or\,\, \mathcal{D}=C\Omega.
\end{equation}
Above and from now on we shall assume fulfilled the following requirements for the function $f(x)$, when defined for  $x\in\mathcal{D}$:
\begin{equation}
\label{1.2}
a)\, f\in L^{\infty}(\mathcal{D}),\, \mathcal{D}=\Omega ;\, b)\,  f\in L^{\infty}(\mathcal{D})\cap L_1(\mathcal{D}),\, \mathcal{D}=C\Omega.
\end{equation}
 It is not difficult to establish that the given in (\ref{1.1}) potential $u_{\alpha ,f}(x)$  is a continuous function in $\mathcal{D}\cup\Gamma$,  therefore the trace
  $\varphi_{\alpha,f}(x)$, $x\in\Gamma$, $\varphi_{\alpha,f}:=u_{\alpha,f}|_{\Gamma}$ is continuous on $\Gamma$. And via the problem of Dirichlet, we found that $u_{\alpha,f}$  is a solution of the equation $(-\Delta)^{\alpha/2} u|_{\mathcal{D}}=f$, with $ u|_{\Gamma}=\varphi_{\alpha,f}$.  The next step is to seek solutions in $S'$  with arbitrary prescribed data, assumed in  $L_2(\Gamma)$. Going to this direction, suppose $u\in S'$   is a solution of the above equation. Then $(-\Delta)^{\alpha/2}[u-u_{\alpha,f}]=0$  in $\Omega\cup C\Omega$  and therefore $L_{\alpha,f}(x):=(-\Delta)^{\alpha/2}[u-u_{\alpha,f}](x)$   is a distribution supported on the surface $\Gamma$. We shall deal here with the case: $L_{\alpha,f}(x)\equiv\delta_{\Gamma}[g](x)$, $x\in\mathbb{R}^3$, i.e. we are interesting in solutions $u\in S'$  of the equation $(-\Delta)^{\alpha/2}u=f^0[f]$ on $\Omega\cup C\Omega$, satisfying the condition 
\begin{equation}
\label{1.3}
(-\Delta)^{\alpha/2}[u-u_{\alpha,f}]=\delta_{\Gamma}[g]\,\, in \,\,S'.
\end{equation}
Above $\delta_{\Gamma}[g]$  is the supported on $\Gamma$  delta function of Dirac, with a density function $g=g(x)\in L_2(\Gamma)$. (As known, e.g. \cite{Hermander}, the action ( $\delta_{\Gamma}[g], \, \phi$) of $\delta_{\Gamma}[g]$  on an arbitrary $\phi\in S$ is defined by the next surface integral, $(\delta_{\Gamma}[g],\phi):=\int_{\Gamma}g(y)\phi(y)ds_y$.)
The important partial case  $f=0$ ( $u_{\alpha,f}=0$) concerns these distributions $w\in S'$, solving the equation below (for $g$ varying in $L_2(\Gamma)$):
\begin{equation}
\label{1.3.0}
(-\Delta)^{\alpha/2}w=\delta_{\Gamma}[g]\,\, in \,\,S'.
\end{equation}
We call BF harmonic (basic fractional harmonic) functions in $\mathbb{R}^3$  such solutions $w$ , and the family $S'_{\alpha,f}$ -- of all solutions $u\in S'$ to (\ref{1.3}) (with $g$ varying in $L_2(\Gamma)$), can be called GS (global solutions) family.

{\bf Remarks:} 1) Clearly for each two distributions $u_1,\, u_2\in S'_{\alpha,f}$  the difference $u_2-u_1$  is a BF harmonic function. \\
2) A possibly larger class of solutions $u\in S'$ to the equation $(-\Delta)^{\alpha/2}u=f^0[f]$ could be expected in the case $L_{\alpha,f}=\delta_{\Gamma}[g_0]+\partial_n\delta_{\Gamma}[g_1]$, where $g_0,\, g_1\in L_2(\Gamma)$ and $\partial_n\delta_{\Gamma}[g](x)$ is the normal to $\Gamma$ derivative of $\delta_{\Gamma}[g]$ at the point $x\in\Gamma$. 

It is not difficult to get a structure description of the family  $S'_{\alpha,f}$. After using the Fourier transform to the relation (\ref{1.3}) one can directly resolve (\ref{1.3}) regarding   $u-u_{\alpha,f}$ and find this manner the next general formula:
\begin{equation}
\label{1.4}
u=\delta_{\Gamma}[g]*\frac{c_{3,\alpha}}{|.|^{3-\alpha}}+u_{\alpha,f}\,\, in \,\,S'.
\end{equation}
The above convolution $\delta_{\Gamma}[g]*\frac{c_{3,\alpha}}{|.|^{3-\alpha}}$ evidently introduces the single layer Riesz potential $v_{\alpha ,g}(x):=\int_{\Gamma}\ds{\frac{c_{3,\alpha}g(y)dy}{|x-y|^{3-\alpha}}}$, ($x\in \mathbb{R}^3$) which possesses well defined direct values $\psi_{\alpha,g}(x)$, $x\in \Gamma$, $\psi_{\alpha,g}:=v_{\alpha,g}|_{\Gamma}$, with $\psi_{\alpha,g}\in L_2(\Gamma)$. This holds because the integral operator $B_{\alpha,\Gamma}[g]:=\psi_{\alpha,\Gamma}$, $g\in  L_2(\Gamma)$, has a weak singularity ( $3-\alpha<2$) and, according to the known classical theory (e.g. \cite{Mikhlin}, \cite{Zabreyko}, \cite{Landkof}), the map 
$B_{\alpha,\Gamma}\, :\, L_2(\Gamma)\rightarrow L_2(\Gamma)$  is a bounded linear operator. As seen from (\ref{1.4}), each solution $u\in S'_{\alpha ,f}$  has an $L_2$  trace on   $\Gamma$ (for $f$  satisfying (\ref{1.2})). Let us also check that the term $\ds{\delta_{\Gamma}[g]*\frac{c_{3,\alpha}}{|.|^{3-\alpha}}}$ is a BF harmonic function: 
\begin{eqnarray*}
\left((-\Delta)^{\alpha/2}\delta_{\Gamma}[g]*\frac{c_{3,\alpha}}{|.|^{3-\alpha}}\right) & = &\delta_{\Gamma}[g]*(-\Delta)^{\alpha/2}\frac{c_{3,\alpha}}{|.|^{3-\alpha}}\\
& = & \delta_{\Gamma}[g]*\delta=\delta_{\Gamma}[g].
\end{eqnarray*}
Concluding the above results, we already found that $(-\Delta)^{\alpha/2}u_{\alpha,f}=f^0[f]$ and 
$$\left((-\Delta)^{\alpha/2}\delta_{\Gamma}[g]*\frac{c_{3,\alpha}}{|.|^{3-\alpha}}\right)=\delta_{\Gamma}[g]$$
 (both in $S'$), i. e.  $(-\Delta)^{\alpha/2}u=\delta_{\Gamma}[g]+f^0[f]$ in $S'$ (for each $g\in L_2(\Gamma)$), $u$ given by (\ref{1.4}), and $u|_{\Gamma}=\psi_{\alpha,g}+\varphi_{\alpha,f}$ ($f$ satisfying (\ref{1.2})).

Now the final question is whether a possibly unique $g\in L_2(\Gamma)$  can be determined, corresponding to  $\varphi$,  for arbitrary $\varphi$  in a suitable subspace of $L_2(\Gamma)$. Then, by the formula (\ref{1.4}) we could get a solution   of the basic problem:
\begin{equation}
\label{1.5}
a)\, (-\Delta)^{\alpha/2}u|_{\mathcal D}=f\, ,\,  b)\, u|_{\Gamma}=\varphi .
\end{equation}
And this solution is expected as the unique one in the family $S'_{\alpha,f}$. We give a positive answer of this question by introducing in Section 3 (below) the subspace $H^1_{\alpha}\subset L_2(\Gamma)$ (coincident with the map image of $B_{\alpha,\Gamma}[L_2(\Gamma)]$), and finding then a unique $g\in L_2(\Gamma)$  such that $B_{\alpha,\Gamma}[g]=\varphi-\varphi_{\alpha,f}$, for $\varphi\in L_2(\Gamma)$  : $\varphi-\varphi_{\alpha,f}\in H^1_{\alpha}(\Gamma)$. The key instrument for obtaining the answer is contained in the properties of the boundary operator $B_{\alpha,\Gamma}$, analyzed primarily in the next Section 2. 

\section{The zero kernel of the boundary integral operator}
It turns out that the kernel of the operator $B_{\alpha,\Gamma}$  (acting from $L_2(\Gamma)$  into $L_2(\Gamma)$) consists only of the zero element  $g=0$, i.e. the unique solution of the equation $B_{\alpha,\Gamma}[g]=0$  is $g=0$. The key to this very important property lies in a simple but essential relation in the form
$I_{\alpha}({\infty} )=const . J_{\Gamma,\alpha}$,
 where $I_{\alpha}({\infty})=\ds{\lim_{r\to\infty}}I_{\alpha}(r) $
 and the terms $I_{\alpha}(r), \, J_{\Gamma,\alpha}$  present respectively the integrals:
\begin{eqnarray*}
I_{\alpha}(r)=\int_{|\xi|\le r}|\hat{\delta}_{\Gamma}[g]|^2|\xi|^{-\alpha}{d\xi};\\
J_{\Gamma,\alpha}=\int_{\Gamma}g(x)(\overline{\delta}_{\Gamma}*|.|^{\alpha-3})(x)ds_x .
\end{eqnarray*}
Clearly the above relation (when holds) means in particular that the integral $I_{\alpha}(\infty)=\int_{\mathbb{R}^3}|\hat{\delta}_{\Gamma}[g]|^2|\xi|^{-\alpha}{d\xi}$  converges.
(Here $\hat{\delta}_{\Gamma}[g](\xi) $ is the Fourier image of   $\delta_{\Gamma}[g](x)$.) The mentioned equality shall be found as a specific consequence of the well known Parseval equality (e.g. \cite{Hermander}, \cite{Reed}). To this goal we shall begin by considering a complement to Parseval's equality idea.
\begin{prop}
\label{p2.1}
({\bf The boundary Parseval formula.})
The following relation is valid, for each function  $\psi\in C^{\infty}(\mathbb{R}^3)$, with  $\hat{\psi}\in L_1(\mathbb{R}^3)$:
\begin{equation}
\label{2.1}
(2\pi)^3(\delta_{\Gamma}[g],\overline{\psi})=(\hat{\delta}_{\Gamma}[g],\hat{\overline{\psi}}).
\end{equation}
\end{prop}

{\bf Proof:}

Note firstly that  $\overline{\psi}$ is the complex conjugated quantity to $\psi$ and recall that the notation $(\delta_{\Gamma}[g],\overline{\psi})$  expresses the action of ${\delta}_{\Gamma}[g]$, as a distribution in $S'$, on the function $\overline{\psi}$ -- as an arbitrary element of $S$. Thus 
$(\delta_{\Gamma}[g],\overline{\psi})=\int_{\Gamma}g(x)\overline{\psi}(x)ds_x$, and by analogy about the notation  $(\hat{\delta}_{\Gamma}[g],\hat{\overline{\psi}})$, i.e.   
\begin{eqnarray}
\label{2.2}
(\hat{\delta}_{\Gamma}[g],\hat{\overline{\psi}}) & = & \int_{\mathbb{R}^3}\hat{\delta}[g](\xi)\hat{\overline{\psi}}(\xi)d\xi \nonumber\\
& = & \int_{\mathbb{R}^3}\hat{\overline{\psi}}(\xi)\int_{\Gamma}g(x)\exp(-i\langle x,\xi\rangle)ds_xd\xi .
\end{eqnarray}
The proof uses the approximation approach to (\ref{2.1}) following the two step scheme: obtain firstly (\ref{2.1}) with $w\in C_0^{\infty}(\mathbb{R}^3)$  instead of 
${\delta}_{\Gamma}[g]$, $ C_0^{\infty}(\mathbb{R}^3)$  is the space of the compactly supported infinitely smooth functions. And apply afterwards an approximation procedure with $w_n$  ($w_n\in C_0^{\infty}(\mathbb{R}^3)$, $n=1,\, 2,\dots $) tending to ${\delta}_{\Gamma}[g]$, at  $n\rightarrow\infty$. The first step is done in the given lemma.         
\begin{lm}
\label{l2.1}
The next Parseval equality is valid for each pair $w\in C_0^{\infty}(\mathbb{R}^3)$ and  $\psi\in C^{\infty}(\mathbb{R} ^3)$, with $\hat{\psi}\in L_1(\mathbb{R}^3)$:
\begin{equation}
\label{2.3}
(2\pi)^3(w,\overline{\psi})=(\hat{w},\hat{\overline{\psi}}).
\end{equation}
\end{lm}
At the beginning to the proof of (\ref{2.3}), note as above that the notation $(w,\overline{\psi})$  is used in the known distribution ( $S'$) sense, with 
$(w,\overline{\psi})=\int_{\mathbb{R}^3}w(x)\overline{\psi}(x)dx$, and by analogy about the notation $(\hat{w},\hat{\overline{\psi}})$. 
Now, let  us introduce the function $\phi_0(x)\in C_{0}^{\infty}$: $\phi_0(x)=\phi_0(|x|)$, $1\ge\phi_0(x)\ge 0 $, $\forall x$, $\phi_0(x)\equiv 1$  for  $|x|\le r^0/2$, 
$\phi_0(x)\equiv 0$ for $|x|\ge r^0$ , with a fixed $r^0>0$  such that $\int_{\mathbb{R}^3}\phi_0(x)dx=1$. By the real parameter $s\in (0,\, 1]$  we will deal with $\phi_0(sx)$  and its Fourier map $\mathbb{F}[\phi_0(s.)](\xi)=\frac{1}{(2\pi s)^3}\hat{\phi }_0(s^{-1}\xi )$. Then the conventional Parseval formula yields the identity $Q(s)=(2\pi)^{-3}\tilde{Q}(s)$, for  
$s\in (0,\, 1]$, where  $Q(s):=(w,\overline{\psi}\phi_0(s.))$ and $\tilde{Q}(s):=(\hat{w},\hat{\overline{\psi}}*(2\pi s)^{-3}\hat{\phi}_0(./s))$. For our goal we have to compare the limit values of $Q(s)$ and $\tilde{Q}(s)$ at  $s\rightarrow 0$. Function $Q(s)$ is actually defined and continuous in $[0,\, 1 ]$, i.e. its limit value ( $s\rightarrow 0$) is  $Q(0)$, while concerning the $\ds{\lim_{s\to 0}\tilde{Q}(s)}$  we need some reworking of  the integral for $\tilde{Q}(s)$. Starting from the initial expression of $\tilde{Q}(s)$ and, applying the linear transform $\theta=s^{-1}(\xi-\eta)$  in the repeated integral (below), we consecutively find the next relations: 
\begin{eqnarray*}
\tilde{Q}(s) & = & \int_{\mathbb{R}^3}\hat{w}(\xi )\hat{\overline{\psi}}*(2\pi s)^{-3}\hat{\phi}_0(s^{-1}.)(\xi )d\xi \\
                  & = &  \int_{\mathbb{R}^3}\hat{w}(\xi)\int_{\mathbb{R}^3}\hat{\overline{\psi}}(\eta)(2\pi s)^{-3}\hat{\phi}_0(s^{-1}(\xi-\eta))d\eta d\xi \\
                   & = &  \int_{\mathbb{R}^3}\hat{\overline{\psi}}(\eta)\int_{\mathbb{R}^3}\hat{w}(\eta+s\theta)\frac{\hat{\phi}_0(\theta)}{(2\pi)^3}d\theta d\eta .
\end{eqnarray*}
The above integral $\int_{\mathbb{R}^3}\hat{w}(\eta+s\theta)\hat{\phi}_0(\theta)d\theta$ is uniformly convergent respectively the parameters $(\eta,\, s)\in K^0\times [0,\, 1]$ , for each compact $K^0\subset\mathbb{R}^3$. (This clearly holds because $\hat{w}(\xi )$  is a bounded function.) Therefore 
$\tilde{F}_w^0(\eta, \, s):=\int_{\mathbb{R}^3}\hat{w}(\eta+s\theta)\hat{\phi}_0(\theta)d\theta$  is a continuous, bounded function in  $\mathbb{R}^3\times[0,\, 1]$, and repeating the same argument (now that $\int_{\mathbb{R}^3}\hat{\overline{\psi}}(\eta )\tilde{F}_w^0(\eta, \, s)d\eta$ is also a uniformly convergent integral) we get that 
$\tilde{f}_w^0(s):=\int_{\mathbb{R}^3}\hat{\overline{\psi}}(\eta )\tilde{F}_w^0(\eta, \, s)d\eta$ is a continuous function in $[0,\, 1]$. However  $\tilde{Q}(s)$ is identical with 
$(2\pi )^{-3}\tilde{f}_w^0(s)$ for $0<s\le1$, and $\ds{\lim_{s\to 0}\tilde{Q}(s)}=\tilde{f}_w^0(0)/(2\pi)^3=(\hat{w},\hat{\overline{\psi}})$, i.e. 
$\ds{\lim_{s\to 0}\tilde{Q}(s)}=(\hat{w},\hat{\overline{\psi}})$. (We have used that  $\frac{1}{(2\pi)^3}\int_{\mathbb{R}^3}\hat{\phi}_0(\theta )d\theta=\phi_0(0)=1$.)   
Thus, letting $s\rightarrow 0$  in the equality $Q(s)=(2\pi)^{-3}\tilde{Q}(s)$, we just obtain the needed formula (\ref{2.3}).

It remains now to perform the approximation step. Suppose $\{w_n(x)\}$, $n=1,\,2,\dots$, is an infinite family of functions $w_n\in C_0^{\infty}(\mathbb{R}^3)$  such that the family of the Fourier maps $\{w_n\}$  is uniformly bounded and $\lim_{n\to\infty}w_n=\delta_{\Gamma}[g]$ (in $S'$). An easy direct construction of such a family is given by the convolution 
$w_n:=\delta_{\Gamma}[g]*n^3\phi_0(nx)$. In this case it is well known (and can be easily verified) that $\ds{\lim_{n\to\infty}}w_n=\delta_{\Gamma}[g]$  in $S'$ , and the assumption for an uniformly bounded $\{\hat{w}_n\}$  is directly seen from $\hat{w}_n=\hat{\delta}_{\Gamma}[g]\hat{\phi}_0(./n)$  (clearly  $\hat{\delta}_{\Gamma}[g]$ and 
$\hat{\phi}_0$  are bounded functions). Letting now $n\rightarrow\infty$ in the equality $(2\pi)^3(w_n,\overline{\psi})=(\hat{w}_n,\hat{\overline{\psi}}).$  (see (\ref{2.3})), we respectively get: $\ds{\lim_{n\to\infty}}(w_n,\overline{\psi})=(\delta_{\Gamma}[g],\overline{\psi})$, and  
$\ds{\lim_{n\to\infty}}(\hat{w}_n,\hat{\overline{\psi}})=(\hat{\delta}_{\Gamma}[g],\hat{\overline{\psi}})$, for $\hat{w}_n=\hat{\delta}_{\Gamma}[g]\hat{\phi}_0(./n)$. Here we have taken into account the equality 
$(\hat{w}_n,\hat{\overline{\psi}})=\int_{\mathbb{R}^3}\hat{\delta}_{\Gamma}[g](\xi )\hat{\phi}_0(\xi /n)\hat{\overline{\psi}}(\xi)d\xi $, combined with the estimate 
$|\hat{\delta}_{\Gamma}[g](\xi )\hat{\phi}_0(\xi/n)\hat{\overline{\psi}}(\xi)|\le (mes(\Gamma))^{1/2}||g||_{L_2(\Gamma)}|\hat{\overline{\psi}}(\xi )|$,  $\xi\in\mathbb{R}^3$, and applying then the well known Lebesgue dominated convergence theorem (e.g. \cite{Rudin}), we find: 
$$\lim_{n\to\infty}\int_{\mathbb{R}^3}\hat{\delta}_{\Gamma}[g](\xi )\hat{\phi}_0(\xi /n)\hat{\overline{\psi}}(\xi)d\xi=\int_{\mathbb{R}^3}\hat{\delta}_{\Gamma}[g](\xi)\hat{\overline{\psi}}(\xi)d\xi=(\hat{\delta}_{\Gamma}[g],\hat{\overline{\psi}}).$$
(Above $mes(\Gamma)$  is the measure of $\Gamma$ and  $||g||_{L_2(\Gamma)}$ is the $L_2$ norm of density $g$.) This proves the boundary Parseval formula (\ref{2.1}).  

Below we add a consequence of (\ref{2.1}), useful for the basic result in this section.
\begin{Cor}
\label{c2.1}
For each $\phi\in C_0^{\infty}(\mathbb{R}^3)$ the next Parseval type relation holds, with $\phi_\mathbb{F}=\mathbb{F}^{-1}[\overline{\phi}]$:
\begin{equation}
\label{2.4}
\int_{\mathbb{R}^3}|\hat{\delta}_{\Gamma}[g](\xi)|^2 .\frac{\phi (\xi)}{|\xi|^{\alpha}}d\xi=(2\pi)^3\int_{\Gamma}g(x)\left(\overline{\delta}_{\Gamma}[g]*\frac{c_{3,\alpha}}{|.|^{3-\alpha}}*\overline{\phi}_{\mathbb{F}}\right)(x)ds_x .
\end{equation}
\end{Cor}

{\bf Proof :} 

Let us set  $\psi_g(x)=\left({\delta}_{\Gamma}[g]*\frac{c_{3,\alpha}}{|.|^{3-\alpha}}*\phi_\mathbb{F}\right)(x)$.Then $\hat{\overline{\psi}}_g(x)=\hat{\overline{\delta}}_{\Gamma}[g](\xi )\frac{\phi (\xi)}{|\xi|^{\alpha}}$ and  $(\hat{\delta}_{\Gamma}[g],\hat{\overline{\psi}}_g)=\int_{\mathbb{R}^3}|\hat{\delta}_{\Gamma}[g](\xi)|^2 .\frac{\phi (\xi)}{|\xi|^{\alpha}}d\xi$. In addition  $({\delta}_{\Gamma}[g],{\overline{\psi}}_g)$ evidently equals to the right hand integral above. Clearly, it is not difficult to check the two assumptions regarding $\psi_g$ : as a first, it is directly seen that $\hat{\psi}_g\in L_1(\mathbb{R}^3)$, secondly, from $\psi_g=\psi_{g,\alpha}*\phi_F$, where  
$\psi_{g,\alpha}:=\delta_{\Gamma}[g]*\frac{c_{3,\alpha}}{|.|^{3-\alpha}}$ is a distribution in $L_1^{loc}(\mathbb{R}^3)$, the validation whether  $\psi_g\in C^{\infty}(\mathbb{R}^3)$ gets obvious. Thus the proof of (\ref{2.4}) follows directly from (\ref{2.1}).

Now the basic result in Section 2 can be presented.

\begin{thm}
\label{th2.1}
({\bf The kernel of $B_{\alpha,\Gamma}$}.)

The zero is not an eigen value of the boundary integral operator  
$$B_{\alpha ,\Gamma}\, : \, L_2(\Gamma)\rightarrow L_2(\Gamma),$$
 i.e. the only solution of the equation $B_{\alpha,\Gamma}[g]=0$  is  $g=0$.
\end{thm}
{\bf Proof :} 

Using (\ref{2.4}) with  $\phi(\xi )\equiv \phi_0(\sigma\xi)$, $\xi\in \mathbb{R}^3$, where   $\sigma\in (0,\, 1]$ is a real parameter, we get the formula:
\begin{equation}
\label{2.5}
\int_{\mathbb{R}^3}|\hat{\delta}_{\Gamma}[g](\xi)|^2 .\frac{\phi _0(\sigma\xi)}{|\xi|^{\alpha}}d\xi=\int_{\Gamma}g(x)\left(\overline{\delta}_{\Gamma}[g]*\frac{c_{3,\alpha}}{|.|^{3-\alpha}}*{\sigma}^{-3}\hat{\phi}_{0}(./\sigma)\right)(x)ds_x .
\end{equation}
(Note here hat $\hat{\phi}_{0}$ is a real valued function.) At the auxiliary assumption for $g$ as a continuous function on $\Gamma$ we will firstly analyze the limit values of the integrals above, for $\sigma\rightarrow 0$. Clearly the limit expression of (\ref{2.5}) is expected in the form:
\begin{equation}
\label{2.6}
\int_{\mathbb{R}^3}\frac{|\hat{\delta}_{\Gamma}[g](\xi)|^2}{|\xi|^{\alpha}}d\xi=(2\pi)^3\int_{\Gamma}g(x)\left(\overline{\delta}_{\Gamma}[g]*\frac{c_{3,\alpha}}{|.|^{3-\alpha}}\right)(x)ds_x .
\end{equation}
We shall start with the integral 
$J_{\Gamma,\alpha}^0(\sigma):=\int_{\Gamma}g(x)\left(\overline{\delta}_{\Gamma}[g]*\frac{c_{3,\alpha}}{|.|^{3-\alpha}}*{\sigma}^{-3}\hat{\phi}_{0}(./\sigma)\right)(x)ds_x$.
(The left integral  $I_{\Gamma,\alpha}^0(\sigma)$ in (\ref{2.5}), with 
 $I_{\Gamma,\alpha}^0(\sigma):=\int_{\mathbb{R}^3}|\hat{\delta}_{\Gamma}[g](\xi)|^2 .\frac{\phi_0 (\sigma\xi)}{|\xi|^{\alpha}}d\xi$, will be comment later.) From the simplified expression 
$J_{\Gamma,\alpha}^0(\sigma)=\int_{\Gamma} g(x)J_{\Gamma,\phi}^0(x;\sigma)d s_x $, where

\begin{eqnarray*} 
J_{\Gamma,\phi}^0 (x;\sigma) & := & \left(\overline{\delta}_{\Gamma}[g]*\frac{c_{3,\alpha}}{|.|^{3-\alpha}}*{\sigma}^{-3}\hat{\phi}_0(./\sigma )\right)(x)\\
               & = & \int_{\Gamma}\overline{g}(y) \int_{\mathbb{R}^3}\frac{c_{3,\alpha}\sigma^{-3}\hat{\phi}_0(t\sigma^{-1})}{|x-y-t|^{3-\alpha}}dtds_y,
\end{eqnarray*} 
it is directly seen that  $J_{\Gamma,\alpha}^0 (\sigma)$ can be presented as follows (applying the substitution  $t\sigma^{-1}=\tau$):
\begin{eqnarray}
\label{2.7}
J_{\Gamma,\alpha}^0(\sigma) & = & \int_{\Gamma}g(x)\int_{\Gamma}\overline{g}(y)\int_{\mathbb{R}^3}\frac{c_{3,\alpha}\sigma^{-3}\hat{\phi}_0(t\sigma^{-1})}{|x-y-t|^{3-\alpha}}dtds_yds_x\nonumber\\
& = & \int_{\mathbb{R}^3}\hat{\phi}_0(\tau)\int_{\Gamma}g(x)\int_{\Gamma}\frac{c_{3,\alpha}\overline{g}(y)ds_y}{|x-\sigma\tau -y|^{3-\alpha}}ds_xd\tau.
\end{eqnarray}
Thus, $J_{\Gamma,\alpha}^0(\sigma)=\int_{\mathbb{R}^3}\hat{\phi}_0(\tau)\int_{\Gamma}g(x)F_g(x-\sigma\tau)ds_xd \tau $, with 
$F_g(\theta):=\int_{\Gamma}\ds{\frac{c_{3,\alpha}\overline{g}(y)ds_y}{|\theta -y|^{3-\alpha}}}$. Note that $F_g(\theta)$ is bounded and continuous function for  
$\theta\in\mathbb{R}^3$, because the given single layer Riesz potential (defining $F_g$) is uniformly convergent regarding  $\theta$, for  $\theta\in K\subset\mathbb{R}^3$,  $K$ an arbitrary fixed compact set containing the closed surface $\Gamma $, under assumption for the continuous surface density  $g$ and the second order regularity of $\Gamma$. This holds by the same arguments well known from the classical potential theory (e.g. \cite{Mikhlin}) of the single layer potential (the case of $\alpha=2$). Next, the found properties of $F_g(\theta)$  yield the automatic conclusion that the function $G(\theta ):=\int_{\Gamma}g(x)F_g(x-\theta)ds_x$ is also bonded and continuous,  $\theta\in\mathbb{R}^3$. Then, again by the mentioned Lebesgue theorem, we see that the integral  $\int_{\mathbb{R}^3}\hat{\phi}_0(\tau)G(\sigma\tau)d\tau$ is uniformly convergent regarding 
$\sigma\in [0,\,1]$, i.e. $J_{\Gamma,\alpha}^0(\sigma)=\int_{\mathbb{R}^3}\hat{\phi}_0(\tau)G(\sigma\tau)d\tau$ is a continuous function in $[0,\,1]$. We get this way:
$\exists \, \ds{\lim_{\sigma\to 0}J_{\Gamma,\alpha}^0(\sigma)}=J_{\Gamma,\alpha}^0(0)$. As clear from (\ref{2.7}), 
$J_{\Gamma,\alpha}^0(0)=\int_{\mathbb{R}^3}\hat{\phi}_0(\tau)d\tau\int_{\Gamma}g(x)\int_{\Gamma}\frac{c_{3,\alpha}\overline{g}(y)ds_y}{|x -y|^{3-\alpha}}ds_x$, i.e. (because of equality $\int_{\mathbb{R}^3}\hat{\phi}_0(\tau)d\tau=(2\pi)^3$) $J_{\Gamma,\alpha}^0(0)=(2\pi)^3\int_{\Gamma}g(x)(\overline{\delta}_{\Gamma}[g]*\frac{c_{3,\alpha}}{|.|^{3-\alpha}})(x)ds_x$) (see the right hand side of (\ref{2.6})). In addition (\ref{2.5}) also yields: $\exists \, \ds{\lim_{\sigma\to 0}I_{\Gamma,\alpha}^0(\sigma)}=J_{\Gamma,\alpha}^0(0)$. On the other hand, from the estimates $I_{\Gamma,\alpha}^0(r^0r^{-1})\le I_{\alpha}(r)\le I_{\Gamma,\alpha}^0(r^0r^{-1}/2)$  we establish that there exists the limit value $I_{\alpha}(\infty):=\ds{\lim_{r\to\infty}I_{\alpha}(r)}$, i.e. the integral  
$\int_{\mathbb{R}^3}|\hat{\delta}_{\Gamma}[g]|^2|\xi|^{-\alpha}d\xi $ converges and its value $I_{\alpha}(\infty)$ equals to  $J_{\Gamma,\alpha}^0(0)$. Thus (\ref{2.6}) is proven. 

Let us look afterwards whether formula (\ref{2.6}) remains valid in the general case  $g\in L_2(\Gamma)$. Actually it will be enough to establish the next inequality:
\begin{equation}
\label{2.8}
I_{\alpha}(r)[g]\le (2\pi)^3\int_{\Gamma}g(x)B_{\alpha,\Gamma}[\overline{g}](x)ds_x,\,g\in L_2(\Gamma).
\end{equation}
Here  $r>0$ is an arbitrary fixed, $I_{\alpha}(r)[g]$ is the previously given integral  $I_{\alpha}(r)$, and for $x\in \Gamma$:  
$B_{\alpha,\Gamma}[\overline{g}](x)\equiv (\overline{\delta}_{\Gamma}[g]*\frac{c_{3,\alpha}}{|.|^{3-\alpha}})(x)$. Note firstly that the integrals $I_{\alpha}(r)[g]$ and 
$\int_{\Gamma}g(x)B_{\alpha,\Gamma}[\overline{g}](x)ds_x$  are correctly defined  $\forall \, g\in L_2(\Gamma)$. Choosing now an arbitrary approximating sequence  
$\{g_n\}\, :\, g_n\rightarrow g $, $n\to\infty$ in  $L_2(\Gamma)$, $g_n$ -- continuous ($\forall \, n=1,\, 2,\, \dots$), we evidently have from (\ref{2.6}) the estimate:
\begin{equation}
\label{2.8,n}
I_{\alpha}(r)[g_n]\le (2\pi)^3\int_{\Gamma}g_n(x)B_{\alpha,\Gamma}[\overline{g}_n](x)ds_x.
\end{equation}
Letting then $n\rightarrow\infty$  in (\ref{2.8,n}), we have preliminary to verify that  $I_{\alpha}(r)[g_n]$ and $\int_{\Gamma}g_n(x)B_{\alpha,\Gamma}[\overline{g}_n](x)ds_x$ respectively tend to $I_{\alpha}(r)[g]$  and $\int_{\Gamma}g(x)B_{\alpha,\Gamma}[\overline{g}](x)ds_x$. Certainly, first of all the below relations evidently hold,
\begin{eqnarray*}
|\hat{\delta}_{\Gamma}[g](\xi)-\hat{\delta}_{\Gamma}[g_n](\xi)| & = & |\int_{\Gamma}[g(x)-g_n(x)]\exp{(-i \langle x,\xi \rangle)}ds_x |\\
           & \le & \left( mes(\Gamma)\right)^{1/2}||g-g_n||_{L_2(\Gamma)},
\end{eqnarray*}
consequently  $|\hat{\delta}_{\Gamma}[g_n](\xi)|$ uniformly tends (at $n\rightarrow \infty$) to $|\hat{\delta}_{\Gamma}[g](\xi)|$, for $|\xi|\le r$, and the same is valid concerning   $|\hat{\delta}_{\Gamma}[g_n](\xi)|^2$ and  $|\hat{\delta}_{\Gamma}[g](\xi)|^2$. Therefore  $\ds{\lim_{n\to\infty}I_{\alpha}(r)[g_n]}=I_{\alpha}(r)[g]$. On the other hand it is not difficult to find:
\begin{eqnarray*}
|\int_{\Gamma}\left(g(x)B_{\alpha,\Gamma}[\overline{g}](x)-g_n(x)B_{\alpha,\Gamma}[\overline{g}_n](x)\right)ds_x| & \le & ||g-g_n||_{L_2(\Gamma)}.||B_{\alpha,\Gamma}[\overline{g}]||_{L_2(\Gamma)}\\
& + & ||g_n||_{L_2(\Gamma)}.||B_{\alpha,\Gamma}||.||g-g_n||_{L_2(\Gamma)},
\end{eqnarray*}
($||B_{\alpha,\Gamma}||$  is the norm of the operator $B_{\alpha,\Gamma}$); i.e. the integral 
$\int_{\Gamma}g_n (x)B_{\alpha,\Gamma}[\overline{g}_n](x)ds_x$ tends to  $\int_{\Gamma}g(x)B_{\alpha,\Gamma}[\overline{g}](x)ds_x$ ($n\rightarrow\infty$). 
Thus the estimate (\ref{2.8}) is proved, and observing that the function of  $r$  $I_{\alpha}(r)[g]$is monotone increasing and bounded (because of (\ref{2.8})) we conclude that the integral $\int_{\mathbb{R}^3}|\hat{\delta}_{\Gamma}[g]|^2|\xi|^{-\alpha}d\xi=I_{\alpha}(\infty)[g]:=\ds{\lim_{r\to\infty}I_{\alpha}(r)[g]}$ converges and the following inequality is fulfilled:
\begin{equation}
\label{2.9}
\int_{\mathbb{R}^3}|\hat{\delta}_{\Gamma}[g]|^2|\xi|^{-\alpha}d\xi\le (2\pi)^3\int_{\Gamma}g(x)B_{\alpha,\Gamma}[\overline{g}](x)ds_x,\, g\in L_2(\Gamma).
\end{equation}
Finally, when $B_{\alpha,\Gamma}[{g}]=0$  evidently  $B_{\alpha,\Gamma}[\overline{g}]=0$ as well, and (\ref{2.9}) shows that  $\hat{\delta}_{\Gamma}[g]=0$, consequently  
${\delta}_{\Gamma}[g]=0$ which automatic yields  $g=0$. This proves the theorem. 

\section{Main results}

The found property of the operator  $B_{\alpha,\Gamma}$ is certainly of an essential importance in our approach for solving the problem of Dirichlet. It is in a direct relation with the well known Hilbert-Schmidt theorem (e.g. \cite{Rudin}, \cite{Reed}) and, as a first step below, we recall a selected formulation of this theorem. (For the proof we refer to the known literature on Functional Analysis.) 
\begin{thm}
\label{th3.0}
({\bf Hilbert-Schmidt})

Let $B\, :\, H\rightarrow H$  be a bounded, compact and symmetrical linear operator in the Hilbert space  $H$, with $h=0$ as the unique solution of the equation $Bh=0$ , $h$ varying in $H$. Then there exists a complete orthogonal system  $\{h_j\}\subset H$,  $||h_j||=1$,  $j=1,2,\dots $, of eigen elements to  $B$, with a corresponding set of  (real) eigen values 
$\{\lambda_j\}$, such that the following expression holds,  $\forall \, h\in H$:

{\bf(H-S)}, \, $h=\ds{\sum_{j=1}^{\infty}\sigma_jh_j}$, $\sigma_j =\langle h,h_j\rangle$.

Here $\langle .\, ,\, .\rangle$  is the scalar product in  $H$ (and $||h||=\langle h,h\rangle $  is the norm of $h$).
\end{thm}

Preparing to apply Theorem \ref{th3.0} concerning the operator  $B_{\alpha,\Gamma}$, we will start with the next two initial properties -- the first one follows from the classical theory of the weakly singular integral equations, and the second -- from Theorem \ref{th3.0}.

(i*) The integral operator  $B_{\alpha,\Gamma}\, :\, L_2(\Gamma)\rightarrow L_2(\Gamma)$, with 
$$B_{\alpha,\Gamma}[g](x):=\ds{\int_{\Gamma}c_{3,\alpha}g(y)|x-y|^{\alpha-3}ds_y},$$ 
$x\in \Gamma $, $g\in L_2(\Gamma)$, is bounded, compact and symmetrical.

(ii*) Each function  $\mu\in L_2(\Gamma)$ can be uniquely expressed by the below given decomposition formula

\begin{equation}
\label{3.1}
\mu=\ds{\sum_{k=1}^{\infty}}\gamma_k\zeta_{k,\alpha},\, in\, L_2(\Gamma),
\end{equation}
where  $\{\zeta_{k,\alpha}\}$ is the complete orthogonal system of eigen functions for $B_{\alpha,\Gamma}$  and  $\gamma_k$ are the Fourier coefficients of $\mu $, 
$\gamma_k :=\int_{\Gamma}\mu(x)\zeta_{k,\alpha}(x)ds_x $.
In our basic result below we shall use the already mentioned subspace  $H_{\alpha}^1(\Gamma)\subset L_2(\Gamma)$.

{\bf Definition 3.1}
{ \it Let us set \\ 
$H_{\alpha}^1(\Gamma):=\{\varphi\in L_2(\Gamma), \varphi=\ds{\sum_{k=1}^{\infty}\tau_k\zeta_{k,\alpha}} : \ds{\sum_{k=1}^{\infty}\tau_k^2\lambda_{k,\alpha}^{-2}}<+\infty\} $,  where $\lambda_{k,\alpha}$ are the eigenvalues of $B_{\alpha,\Gamma}$. The scalar product $\langle\varphi,\psi\rangle_{1,\alpha}$ in $H_{\alpha}^1(\Gamma)$ is defined by the sum  $\ds{\sum_{k=1}^{\infty}\tau_k}\overline{\theta}_k(1+\lambda_{k,\alpha}^{-2})$, for $\varphi , \, \psi\in H_{\alpha}^1(\Gamma)$: $\varphi=\ds{\sum_{k=1}^{\infty}\tau_k\zeta_{k,\alpha}}$, $\psi =\ds{\sum_{k=1}^{\infty}\theta_k\zeta_{k,\alpha}}$.}

Note that the inverse operator  $B_{\alpha,\Gamma}^{-1}$ of $B_{\alpha,\Gamma}$ is correctly defined on  $H_{\alpha}^1(\Gamma)$, by the evident rule  
$B_{\alpha,\Gamma}^{-1}[\varphi ]:=\ds{\sum_{k=1}^{\infty}\tau_k\lambda_{k,\alpha}^{-1}\zeta_{k,\alpha}}$, for
 $\varphi=\ds{\sum_{k=1}^{\infty}\tau_k\zeta_{k,\alpha}}$, $\varphi\in H_{\alpha}^1(\Gamma)$.Thus 
 $B_{\alpha,\Gamma}^{-1}:H_{\alpha}^1(\Gamma)\rightarrow L_2(\Gamma)$  is a bounded linear operator.
 
 In the first theorem below, except results on existence, uniqueness and continuous data dependence of solutions, additional ones are also included concerning the asymptotic 
 (at $|x|\rightarrow\infty$  ) and  $L_1^{loc}(\mathbb{R}^3)$ approximation of solutions (by globally defined continuous functions). As a specific moment, the approximation process is uniquely generated by the corresponding boundary one in  $L_2(\Gamma)$. Consider now the central result of our study.

\begin{thm}
\label{th3.1}
Let $f(x)$  be a function, defined on $\mathcal{D}$ and satisfying the assumptions (\ref{1.2}). Then, for each data  $\varphi(x)\in L_2(\Gamma):\, (\varphi-\varphi_{\alpha,f})(x)\in H_{\alpha}^1(\Gamma)$, the problem of Dirichlet (\ref{1.5}) is solvable in $L_1^{loc}(\mathbb{R}^3)$  by the formula
\begin{equation}
\label{3.2}
u(x)=\int_{\Gamma}\frac{c_{3,\alpha}B_{\alpha,\Gamma}^{-1}[\varphi-\varphi_{\alpha,f}](y)ds_y}{|x-y|^{3-\alpha}}
        +\int_{\mathcal{D}}\frac{c_{3,\alpha}f(y)dy}{|x-y|^{3-\alpha}},\, x\in\mathbb{R}^3.
\end{equation}
The above function $u$  is the unique solution of (\ref{1.5}) in the family $S'_{\alpha,f}$, contained in the class  $L_1^{loc}(\mathbb{R}^3)$ and continuous in the two domain components of $\mathbb{R}^3\setminus\Gamma$. Solution (\ref{3.2}) is additionally characterized by the next conventional but essential properties.

(P1) In case of $f(x)$ with a compact support in $\mathcal{D}$, when $\mathcal{D}=C\Omega$, the asymptotic relation below holds for $u(x)$:
\begin{equation}
\label{3.3}
|u(x)|\le \frac{c_0}{|x|^{3-\alpha}}, \, |x|\rightarrow\infty ,
\end{equation}
(i. e. $u(x)=O({1}/{|x|^{3-\alpha}})$ for $|x|\rightarrow\infty$), with constant $c_0$;

(P2) A property for continuous data dependence is valid in the  $L_1^{loc}(\mathbb{R}^3)$ sense, expressed by the following assertion: given two systems of data, $\{f_1,f_2\}$  -- satisfying (\ref{1.2}) and  $\{\varphi_1,\varphi_2\}\subset L_2(\Gamma)$, where  $(\Delta_f \varphi)_i:=\varphi_i-{\varphi_{\alpha,f}}_i\in H_{\alpha}^1$, $i=1,\, 2$, there exist constants 
$C_K^0$, $C_K^{*}$ so that the difference $u_2-u_1$ of the solutions, corresponding to the above data, satisfies an estimate in the form:
\begin{equation}
\label{3.4}
||u_2-u_1||_{L_1(K)}\le C_K^0 ||(\Delta_f \varphi)_2-(\Delta_f \varphi)_1||_{H_{\alpha}^1(\Gamma)}+C_K^{*}||f_2-f_1||_{L_1(\mathcal{D})},
\end{equation}
for an arbitrary chosen compact  $K\subset \mathbb{R}^3$;

(P3) Each approximating system  $\{\psi_n\}\subset H_{\alpha}^1(\Gamma)$, $\ds{\lim_{n\to\infty}\psi_n}= g_f[\varphi ]$ in  $L_2(\Gamma)$ \\
(where  $g_f[\varphi ]:=B^{-1}_{\alpha,\Gamma}[\varphi-\varphi_{\alpha,f}]$), with continuous functions $\psi_n$, generates an infinite sequence of continuous approximations  $u_n$ to $u$ : $\ds{\lim_{n\to\infty}u_n}=u$   in $L^{loc}_1(\mathbb{R}^3)$. More over, $u_n$ solve the problem (\ref{1.5}) at the boundary condition $u|_{\Gamma}=\varphi_n$, $\lim_{n\to\infty}\varphi_n=\varphi$ in  $L_2(\Gamma)$, with $\varphi_n:=B_{\alpha,\Gamma}[\psi_n]+\varphi_{\alpha,f}$, and the estimate (\ref{3.5}) (below) is valid for each fixed compact 
$K\subset\mathbb{R}^3$:
\begin{equation}
\label{3.5}
||u-u_n||_{L_1(K)} \le C_K^0||\varphi-\varphi_n||_{H_{\alpha}^1(\Gamma)}.
\end{equation}
\end{thm}
{\bf Proof:}

Recall that the verification whether function  $u(x)$ from (\ref{3.2}) satisfies the equation \\
$(-\Delta)^{\alpha/2}u|_{\mathcal{D}}=f$ has been actually done in Section 1: by the notations $v_{\alpha,g}(x)$,  $u_{\alpha,f}(x)$ -- respectively for the already introduced single layer and volume Riesz potential, with $g=g_f[\varphi]$, formula (\ref{3.2}) is rewritten as $u=v_{\alpha,g}+u_{\alpha,f}$, where $v_{\alpha,g}$ is a  BF harmonic function, while  $(-\Delta)^{\alpha/2}u_{\alpha,f}=f^0[f]$ (in $S'$  ). And for 
$(-\Delta)^{\alpha/2}u$  we get $(-\Delta)^{\alpha/2}u=\delta_{\Gamma}[g]+f^0[f]$  (in $S'$), which evidently means that  $(-\Delta)^{\alpha/2}u|_{\mathcal{D}}=f$. 
Next, for  $x\in \Gamma$ we have: 
$u|_{\Gamma}=v_{\alpha,g}|_{\Gamma}+\varphi_{\alpha,f}=B_{\alpha,\Gamma}[B_{\alpha,\Gamma}^{-1}[\varphi-\varphi_{\alpha,f}]]+\varphi_{\alpha,f}=\varphi$. 
Thus the existence assertion is proved (i.e.  $u$ is a solution of the problem (\ref{1.5}) in $S'_{\alpha,f}$). For the uniqueness of solution (\ref{3.2}) in  $S'_{\alpha,f}$, assuming existence of two ones, $u_1,\, u_2\in S'_{\alpha,f}$  which satisfy (\ref{1.5}) (with identical data $\varphi$, $f$ ), it is directly seen that the difference  $U=u_2-u_1$ is a BF harmonic function, i.e. $(-\Delta)^{\alpha/2}U=\delta_{\Gamma}[g]$  in $S'$, with a density $g\in L_2(\Gamma)$. To resolve this equation regarding $U$ (recall the analogous comments about (\ref{1.3})) we have evidently to act by the operation   $\ds{\frac{c_{3,\alpha}}{|.|^{3-\alpha}}*}$, finding thus the expression 
$U(x)=(\delta_{\Gamma}[g]*\ds{\frac{c_{3,\alpha}}{|.|^{3-\alpha}}})(x)$, $x\in \mathbb{R}^3$. Restricted on $\Gamma$ it yields: $U|_{\Gamma}=B_{\alpha,\Gamma}[g]$, 
i.e. $B_{\alpha,\Gamma}[g]=0$, therefore $g=0$ (Theorem \ref{th2.1}), and 
(from  $U=(\delta_{\Gamma}[g]*\ds{\frac{c_{3,\alpha}}{|.|^{3-\alpha}}})$ in $S'$)  $U(x)=0$, $x\in\mathbb{R}^3$. Next, looking at formula (\ref{3.2})
 (i.e. $u=v_{\alpha,g}+u_{\alpha,f}$), it is directly seen that $v_{\alpha,g},\, u_{\alpha,f}\in L_1^{loc}(\mathbb{R}^3) $, and the same for $u(x)$. More over, as in the proof of (\ref{3.4}) (below), it follows the estimate
 \begin{equation}
\label{3.6}
||u||_{L_1(K)} \le C^{0}_K||\varphi-\varphi_{\alpha,f}||_{H_{\alpha}^1(\Gamma)}+C^{*}_K||f||_{L_1(\mathcal{D})}.
\end{equation}
(Here $K\subset\mathbb{R}^3$ is an arbitrary fixed compact, and a choice of constants  $C^{0}_K$, $C^{*}_K$ shall be given concerning (\ref{3.4})). 
 The property $u(x)\in C^0(\mathbb{R}^3\setminus\Gamma)$ ($C^0$-- the space of the continuous functions) in both the cases 
 $\mathcal{D}=\Omega$ and $\mathcal{D}=C\Omega$ is also an automatic consequence from the clear relations $v_{\alpha,g}\in C^0(K)$,  $u_{\alpha,f}\in C^0(K)$, valid for each compact $K\subset\mathbb{R}^3\setminus\Gamma$.

Consider now the properties (P1) -- (P3).The asymptotic relation (\ref{3.3}) is actually evident (as a slight consequence of the standard inequality $|x-y|\ge|x|-|y|>0$, valid at 
$|x|\rightarrow\infty$ and $y$ varying in a compact). For the proof of (\ref{3.4}) let us firstly rewrite (\ref{3.2}) with $u_2-u_1$, $
\varphi_2-\varphi_{\alpha,f_2}-(\varphi_1-\varphi_{\alpha,f_1})$,  $f_2-f_1$, respectively instead of $u$, $\varphi-\varphi_{\alpha,f}$, $f$. For the sake of convenience we will use below the notations $\Delta_f\varphi=\varphi-\varphi_{\alpha,f}$, ($(\Delta_f\varphi)_i=\varphi_i-\varphi_{\alpha,f_i}$, $i=1,2 $). After integration of $|u_2-u_1|$  on a compact $K\subset\mathbb{R}^3$ it easily follows:
\begin{eqnarray*}
|u_2-u_1|_{L_1(K)} & \le & c_{3,\alpha}\int_{\Gamma}W_K(y)|B_{\alpha,\Gamma}^{-1}[(\Delta_f\varphi)_2-(\Delta_f\varphi)_1](y)|ds_y\\
                   & + & c_{3,\alpha}\int_{\mathcal{D}}W_K(y)|f_2(y)-f_1(y)|dy,\\
W_K(y)& := & \int_K|x-y|^{\alpha-3}dx,\, y\in\mathbb{R}^3.
\end{eqnarray*}
In order to rework suitably the above inequality we shall take into account the estimate  
$$||c_{3,\alpha}B_{\alpha,\Gamma}^{-1}[(\Delta_f\varphi)_2-(\Delta_f\varphi)_1]||_{L_2(\Gamma)}\le b_{\alpha, \Gamma}^{*}||(\Delta_f\varphi)_2-(\Delta_f\varphi)_1||_{H_{\alpha}^1(\Gamma)},$$
 where $b_{\alpha, \Gamma}^{*}$ is the norm of the operator $c_{3,\alpha}B_{\alpha,\Gamma}^{-1}$. This way we come to the next relation for the difference $u_2-u_1$:
\begin{eqnarray}
\label{3.7}
||u_2-u_1||_{L_1(K)} &\le & ||W_K||_{L^{\infty}(\mathbb{R}^3)}\left(b_{\alpha, \Gamma}^{*}\sqrt{mes\Gamma}||(\Delta_f\varphi)_2-(\Delta_f\varphi)_1||_{H_{\alpha}^1(\Gamma)}\right)\nonumber\\
& + & ||W_K||_{L^{\infty}(\mathbb{R}^3)}\left( c_{3,\alpha}||f_2-f_1||_{L_1(\mathcal{D})}\right).
\end{eqnarray}
Afterwards it remains to set: $C_K^0=b_{\alpha, \Gamma}^{*}\sqrt{mes\Gamma}||W_K||_{L^{\infty}(\mathbb{R}^3)}$, 
$C_K^*=|c_{3,\alpha}||W_K||_{L^{\infty}(\mathbb{R}^3)}$. Thus (\ref{3.7}) takes the form of (\ref{3.4}). 

{\bf Remark:} The partial case $f_1=f_2$ could be practically more valuable (then the accent is paid on the boundary data dependence). Now the estimates (\ref{3.7}), (\ref{3.4}) take respectively the forms:
\begin{equation}
\label{3.8.a}
 ||u_2-u_1||_{L_1(K)} \le ||W_K||_{L^{\infty}(\mathbb{R}^3)}b_{\alpha, \Gamma}^{*}\sqrt{mes\Gamma}||\varphi_2-\varphi_1||_{H_{\alpha}^1(\Gamma)};
\end{equation}
\begin{equation}
\label{3.8.b}
||u_2-u_1||_{L_1(K)} \le C_K^0||\varphi_2-\varphi_1||_{H_{\alpha}^1(\Gamma)}.
\end{equation}
Consider finally the proof of (P3). Via the remark above, when the boundary problem (\ref{1.5}) is used in a model, the contour $L_2$ data  $\varphi$ can be preferably changed by suitable continuous approximations  $\{\varphi_n\}$ in order to simplify say a numerical procedure. In our approach the boundary operator pair 
$\{B_{\alpha,\Gamma},B_{\alpha,\Gamma}^{-1}\}$ suggests to seek $\{\varphi_n\}$ by the map $B_{\alpha,\Gamma}[\psi_n]$, given an arbitrary sequence  
$\{\psi_n\}\subset H_{\alpha}^1(\Gamma)$ of continuous $L_2$ approximations to $g_f[\varphi]$. In the framework of problem (\ref{1.5}) (considered now at boundary data 
$\varphi_n$, regarding an unknown solution $u_n$) the basic formula (\ref{3.2}) serves the answer both, for  $\varphi_n$ and  $u_n$, namely 
$\varphi_n=B_{\alpha,\Gamma}[\psi_n]+\varphi_{\alpha,f}$  and $u_n$ as follows:
\begin{equation}
\label{3.9}
u_n(x)=\int_{\Gamma}\frac{c_{3,\alpha}\psi_n(y)ds_y}{|x-y|^{3-\alpha}}+u_{\alpha,f}(x),\, x\in\mathbb{R}^3.
\end{equation}
The property  $u_n\in C^0(\mathbb{R}^3)$ follows (by the integral above) from the continuous assumption for $\psi_n$, and the same for $\varphi_n$. (We have taken again into account that the single layer Riesz potentials possess, at $3-\alpha<2$, the same continuous properties as in the classical case of $3-\alpha=1$.) And the announced estimate (\ref{3.5}) is actually proved by the already shown (\ref{3.8.b}). In a conclusion let us comment how to construct approximating systems $\{\psi_n\}\subset H_{\alpha}^1(\Gamma)$ of continuous functions: introducing an arbitrary system of (real) numbers $\{\tau_k\},\, k=1,2,\dots$: $\ds{\sum_{k=1}^{\infty}\tau_k^2\lambda_{k,\alpha}^{-2}}<\infty $,
we get an element $g$ of the space $H_{\alpha}^1(\Gamma)$,  $g(x):=\ds{\sum_{k=1}^{\infty}\tau_k\zeta_{k,\alpha}(x)}$ and for an obviously convenient approximating (to $g$) sequence we have to take $\psi_n (x):=\ds{\sum_{k=1}^{n}\tau_k\zeta_{k,\alpha}(x)}$, $n=1,2,\dots $. Recall that the eigen functions $\zeta_{k,\alpha}$ of $B_{\alpha,\Gamma}$ are continuous (i.e. $\{\zeta_{k,\alpha}\}\subset C^0(\Gamma)$), according to the known classical theorem for the continuous $L_2$ solutions of weakly singular integral equations (e.g. \cite{Mikhlin}, \cite{Zabreyko}, \cite{Stein}).

Our next result concerns regularity properties of the solution (\ref{3.2}), in the interior of  $\mathbb{R}^3\setminus\Gamma$, as a consequence from these of $f(x)$. We consider below two cases for the regularity of $f(x)$, assumed with a compact support:  $f\in C_0^m(\mathcal{D})$, $m=1,2,\dots$, and 
$f\in L^{\infty}(\mathcal{D})\cap \mathfrak{L}'(\mathcal{D})$, at $f^0[f]\in H^s(\mathbb{R}^3)$, $s>0$. Here, as usually, $C_0^m(\mathcal{D})$ is the space of the functions smooth up to order $m$ in $\mathcal{D}$, with compact supports; $ H^s(\mathbb{R}^3)$ are the known Sobolev classes, related to $\mathbb{R}^3$ (e.g. \cite{Hermander}), 
and $\mathfrak{L}'(\mathcal{D})$  is the space of the Lauren Schwartz distributions, defined on  $\mathcal{D}$ (\cite{Hermander}, \cite{Rudin}, \cite{Reed}), possessing compact supports therein. Clearly the elements of $\mathfrak{L}'(\mathcal{D})$ are automatically extended (on the whole $\mathbb{R}^3$) as zeros out of their supports, presenting thus distributions from $S'(\mathbb{R}^3)$. 

Concerning the conventional regularity of the (\ref{3.2}) solution  $u(x)$ (in the case $f\in C_0^m(\mathcal{D})$) we apply below again  $L_1(K)$ estimates, now related to the partial derivatives  $\partial_x^{\beta}u(x)$. Recall here that $\beta $ is a $3D$ multi index, i.e.  $\beta=(\beta_1,\beta_2,\beta_3)$, with $\beta_i$ (i=1,2,3 ) -- (nonnegative) integers; 
$\partial_x^{\beta}u(x)$ is of order  $k$ ( k=0,1,2,\dots) when $|\beta|=k$, $|\beta|:=\beta_1+\beta_2+\beta_3$, and a function $F(x)$, defined in a domain 
$\tilde{\Omega}\subset \mathbb{R}^3$, belongs to the class  $C^m(\tilde{\Omega})$ when $F$ possesses continuous in $\tilde{\Omega}$ derivatives of each order $k$,  $k\le m$. 
In the case of certain Sobolev regularity for the solution $u(x)$ (assuming $f^0[f]\in H^s(\mathbb{R}^3))$, it is clearly expected to hold $u\in H_{loc}^s(\mathbb{R}^3\setminus\Gamma)$. As known, this inclusion is characterized by the property 
$\theta u\in H^s(\mathbb{R}^3)$, valid for each function  $\theta(x)\in C_0^{\infty}(\mathbb{R}^3\setminus\Gamma)$ (at $\theta u$ automatically extended as zero out of the support of $\theta (x)$). And we shall seek a relevant  $H^s$ estimation of $\theta u$ by the boundary data $\varphi-\varphi_{\alpha,f}$ and $f$.    

For analyzing the $H^s$ properties of  $u(x)$ we will use the following accessory assertion (e.g. \cite{Treves}). (The given proof of the lemma is due to the university lectures \cite{Genchev}.)

\begin{lm}
\label{lm3.1}
The map  ${\bf M}_{\Phi}: {\bf v}\rightarrow \Phi {\bf v}$,  ${\bf v}\in H^s$, with $\Phi (x)\in S$ -- an arbitrary fixed function, is a continuous operator, acting from $H^s$ into itself, for each (fixed) real $s$. (Here  $S=S(\mathbb{R}^3)$ and the same for $H^s$.)  
\end{lm}
{\bf Proof:} As a necessary initial step, recall the very useful representation for the Fourier image $(\hat{\Phi{\bf v}})(\xi)$, $\xi\in\mathbb{R}^3$ , of the (generalized) function 
$(\Phi{\bf v})(x)$:
\begin{equation}
\label{3.10}
(\hat{\Phi{\bf v}})(\xi)=(2\pi )^{-3}(\hat{{\bf v}}*\hat{\Phi})(\xi)=(2\pi )^{-3}\int_{\mathbb{R}^3}\hat{\Phi}(\xi-\eta)\hat{\bf v}(\eta )d\eta .
\end{equation}
From (\ref{3.10}), taking into account the known Peetre inequality (e.g. \cite{Treves}), we get:
\begin{eqnarray*}
|(1+|\xi)|^2)^{s/2}(\hat{\Phi{\bf v}})(\xi)| \le \frac{2^{|s/2|}}{(2\pi)^3}\int_{\mathbb{R}^3}(1+|\xi-\eta|^2)^{|s/2|}|\hat{\Phi}(\xi-\eta)(1+|\eta)|^2)^{s/2}\hat{\bf v}(\eta)|d\eta.
\end{eqnarray*}
Applying afterwards the Young inequality (\cite{Hermander}, \cite{Landkof}, \cite{Stein}) in the integral term above, we find the sought estimate:
\begin{equation}
\label{3.11}
||{\bf \Phi v}||_s\le {\bf C_M}||{\bf v}||_s,\, {\bf v}\in H^s.
\end{equation}
Here $||.||_s$ is the norm in the space $H^s$, $||{\bf \Phi v}||_s:=||(1+|\xi)|^2)^{s/2}(\hat{\Phi{\bf v}})(\xi)||_{L_2(\mathbb{R}^3)}$, and\\
 ${\bf C_M}:=\ds{\frac{2^{|s/2|}}{(2\pi)^3}}||(1+|\xi)|^2)^{|s/2|}(\hat{\Phi})(\xi)||_{L_1(\mathbb{R}^3)}$.
We can now formulate and proof our result concerning the regularity of the solution in (\ref{3.2}).
\begin{thm}
\label{th3.2}
Suppose the free term  $f(x)$ in the equation (\ref{1.5}.a) belongs to some of the spaces $C_0^m(\mathcal{D})$, $L^{\infty}(\mathcal{D})\cap \mathfrak{L}'(\mathcal{D})$ with  $f^0[f]\in H^s$. Then the (\ref{3.2}) solution $u(x)$ has the relevant regularity properties in $\mathbb{R}^3\setminus\Gamma$, satisfying the attached estimates, as follows. 

(I)  When  $f\in C_0^m(\mathcal{D})$, it holds that  $u\in C^m(\mathbb{R}^3\setminus\Gamma)$ and the estimate below is valid, for each compact 
$K\subset\mathbb{R}^3\setminus\Gamma$, $\forall \beta: |\beta |\le m$:
\end{thm}
\begin{eqnarray}
\label{3.12}
||\partial_x^{\beta}u||_{L_1(K)}\le {\bf C}_{K,\beta}^0||\varphi-\varphi_{\alpha,f}||_{H_{\alpha}^1(\Gamma)}+{\bf C}_K^*||\partial_x^{\beta}f||_{L_1(\mathcal{D})};\\
{\bf C}_{K,\beta}^0=b_{\alpha,\Gamma}^*\sqrt{mes\Gamma}||W_{K,\beta}||_{C^0(\Gamma)}, W_{K,\beta}(y):=\int_K\partial_x^{\beta}|x-y|^{\alpha-3}dx.\nonumber
\end{eqnarray}

(II) When $f\in L^{\infty}(\mathcal{D})\cap \mathfrak{L}'(\mathcal{D})$: $f^0[f]\in H^s$, for $1<\alpha<3/2$, the inclusion $u\in H^s_{loc}(\mathbb{R}^3\setminus\Gamma) $ is valid and there exist two constants  ${\bf c}_{\theta,1}$, ${\bf c}_{\theta,2}$ (depending on $\theta$), such that the next estimate is fulfilled,  
$\forall\theta\in C_0^{\infty}(\mathbb{R}^3\setminus\Gamma)$:
\begin{equation}
\label{3.13}
||\theta u||_s\le {\bf c}_{\theta,1}||\varphi-\varphi_{\alpha,f}||_{H_{\alpha}^1(\Gamma)}+{\bf c}_{\theta,2}\left(||f||_{L^{\infty}}+||f||_s\right).
\end{equation}
(Above $L^{\infty}=L^{\infty}(\mathbb{R}^3)$.)

{\bf Proof:} In both the cases (I) and (II) we clearly can conveniently deal with the short version of (\ref{3.2}), i.e. $u=v_{\alpha,g}+u_{\alpha,f}$ 
(with $g=B_{\alpha,\Gamma}^{-1}[\Delta_f\varphi]$). Acting by the operation  $\partial_x^{\beta}$ on the components $v_{\alpha,g}$ and $u_{\alpha,f}$, we respectively find that:
\begin{equation*}
\partial_x^{\beta}v_{\alpha,g}(x)=\int_{\Gamma}c_{3,\alpha}B_{\alpha,\Gamma}^{-1}[\Delta_f\varphi](y)\partial_x^{\beta}|x-y|^{\alpha-3}ds_y, \, and\,
\partial_x^{\beta}u_{\alpha,f}(x)=\left(\partial_x^{\beta}f*\frac{c_{3,\alpha}}{|.|^{3-\alpha}}\right)(x),
\end{equation*}
for $x\in K$ ($K$ is a compact in $\mathbb{R}^3\setminus\Gamma$). It becomes now clear the property $u\in C^m(\mathbb{R}^3\setminus\Gamma)$. In addition the above expression for $\partial_x^{\beta}v_{\alpha,g}$ suggests to introduce the function 
$$W_{K,\beta}(y):=\int_K\partial_x^{\beta}|x-y|^{\alpha-3}dx.$$ 
Afterwards, for proving the estimate (\ref{3.12}) we only have to follow the already used steps, known from the proof of (\ref{3.4}): the constant ${\bf C}_{K,\beta}^0$ in (\ref{3.12}) is evidently the analogous one to ${\bf C}_K^0$ and ${\bf C}_K^*$ is just that from the estimate (\ref{3.4}).

Going to the proof of (II), let us multiply the relation $u=v_{\alpha,g}+u_{\alpha,f}$ by an arbitrary $\theta(x)\in C_0^{\infty}(\mathbb{R}^3\setminus\Gamma)$ and consider next the $H^s$  properties of the terms $\theta v_{\alpha,g}$, $\theta u_{\alpha,f}$. According to the Lemma, for the second product we could get the conclusion that 
$\theta u_{\alpha,f}\in H^s$ if $u_{\alpha,f}\in H^s$. But the former certainly holds for $1<\alpha<3/2$ (under the assumption $f^0[f]\in H^s$):
\begin{equation*}
||u_{\alpha,f}||_s^2=||(1+|\xi)|^2)^{s}|\hat{f}(\xi)|^2|\xi|^{-2\alpha}||_{L_1(\mathbb{R}^3)}\le ||\hat{f}||^2_{C^0(|\xi|\le 1)}\int_{|\xi|\le 1}\frac{2^s}{|\xi|^{2\alpha}}d\xi +||f||^2_s.
\end{equation*}
On the other hand  $\theta v_{\alpha,g}\in C_0^{\infty}(\mathbb{R}^3\setminus\Gamma)$, therefore $u\in H_{loc}^s(\mathbb{R}^3\setminus\Gamma)$. Preparing the final estimate (\ref{3.13}), we will firstly specify the above estimate for  $||u_{\alpha,f}||^2$ -- concerning the term with $ ||\hat{f}||^2_{C^0(|\xi|\le 1)}$ it actually holds that:
\begin{equation*}
||\hat{f}||^2_{C^0(|\xi|\le 1)}\int_{|\xi|\le 1}\frac{2^s}{|\xi|^{2\alpha}}d\xi \le\frac{2^{s+2}}{3-2\alpha}\pi mes^2K_f^0||f||^2_{L^{\infty}}.
\end{equation*}
Here $K_f^0=supp[f]$  (the supporter of  $f$) and $L^{\infty}=L^{\infty}(\mathbb{R}^3)$. Consequently  $||u_{\alpha,f}||_s$ satisfies the inequality
\begin{equation}
\label{3.14}
||u_{\alpha,f}||_s\le\left(1+2^{1+s/2}\frac{\sqrt{\pi}}{\sqrt{3-2\alpha}}mes K_f^0 \right)\left(||f||_{L^{\infty}}+||f||_s\right).
\end{equation}
Now by the Lemma \ref{lm3.1} we can easily estimate the product $\theta u_{\alpha,f}$:
\begin{equation}
\label{3.15}
||\theta u_{\alpha,f}||_s\le{\bf C}_{{\bf M},\theta}{\bf C}_f^0\left(||f||_{L^{\infty}}+||f||_s\right).
\end{equation}
(For ${\bf C}_{{\bf M},\theta}$, ${\bf C}_f^0$  we respectively have: ${\bf C}_{{\bf M},\theta}:=2^{|s|/2}(2\pi)^{-3}||(1+|\xi)|^2)^{|s|/2}\hat{\theta}(\xi)||_{L_1}$, with 
$L_1=L_1(\mathbb{R}^3)$, and ${\bf C}_f^0=(1+2^{1+s/2}\ds{\frac{\sqrt{\pi}}{\sqrt{3-2\alpha}}}mes K_f^0 $.)

 It remains then to estimate the product  $\theta v_{\alpha,g}$ ($g=B_{\alpha,\Gamma}^{-1}[\Delta_f,\varphi]$). In order to express conveniently the impact of the boundary data we shall deal with the norm $||\theta v_{\alpha,g}||_{[s]+1}$ (using that $||\theta v_{\alpha,g}||_{[s]}\le ||\theta v_{\alpha,g}||_{[s]+1}$, where  $[s]$ is the integer part of $s$). As known, $||\theta v_{\alpha,g}||_{[s]+1}^2$ can be expressed taking the sum of addends like $||\partial^{\beta}_x\theta v_{\alpha,g}||^2_{L_2}$, where\\ 
 $\partial^{\beta}_x(\theta v_{\alpha,g})(x)=\int_{\Gamma}c_{3,\alpha}B_{\alpha,\Gamma}^{-1}[\Delta_f\varphi](y)\partial^{\beta}_x(\theta v_{\alpha,g}(x)|x-y|^{\alpha-3})ds_y$.
The Cauchy-Schwartz inequality now yields:
\begin{equation}
\label{3.16}
|\partial^{\beta}_x\theta v_{\alpha,g}|^2 \le ||c_{3,\alpha}B_{\alpha,\Gamma}^{-1}[\Delta_f \varphi]||_{L_2(\Gamma)}^2\int_{\Gamma}|\partial^{\beta}_x(\theta (x)|x-y|^{\alpha-3}|^2ds_y.
\end{equation}
Summarizing above on all  $\beta: |\beta|=k$, for $k=0,1,\dots [s]+1$, and taking an integration $\int_{\mathbb{R}^3}|...|^2 dx$ on the relevant terms, we obtain:
\begin{equation}
\label{3.17}
||\theta v_{\alpha,g}||_{[s]+1}^2 \le ||c_{3,\alpha}B_{\alpha,\Gamma}^{-1}[\Delta_f \varphi]||_{L_2(\Gamma)}^2\int_{\Gamma}||\theta |.-y|^{\alpha-3}||^2_{[s]+1}ds_y.
\end{equation}
By the notation  $W_{\alpha, [s]+1}[\theta ](y):=||\theta |.-y|^{\alpha-3}||_{[s]+1} $ (\ref{3.17}) can be evidently rearranged in the next form:
\begin{equation}
\label{3.18}
||\theta v_{\alpha,g}||_{[s]+1}\le b_{\alpha,\Gamma}^*||W_{\alpha, [s]+1}[\theta ]||_{L_2(\Gamma)}||\varphi-\varphi_{\alpha,f}||_{H^1_{\alpha}(\Gamma)}.
\end{equation}
Finally, from the initial inequality $||\theta u||_s\le ||\theta v_{\alpha,g}||_{[s]+1}+||\theta u_{\alpha,f}||_{s}$, and the sum of (\ref{3.15}), (\ref{3.18}) we get the expected estimate (\ref{3.13}), with ${\bf c}_{\theta,1}=b_{\alpha,\Gamma}^*||W_{\alpha, [s]+1}[\theta ]||_{L_2(\Gamma)}$, ${\bf c}_{\theta,2}={\bf C}_{{\bf M},\theta}{\bf C}_f^0$. Thus the theorem is proved.

\vspace{5ex}

\noindent
{\Large \bf Appendix}
\appendix
 \section{The Dirichlet problem for $1D$ equations}
We will comment here the problem of Dirichlet for the $1D$ equations $\ds{(-\Delta)^{\alpha/2}u|_{(l_0,+\infty)}=f(x)}$, with   assumed continuous in $(l_0,\, +\infty)$  and (for the sake of simplicity) $f(x)$ vanishing out of a compact subinterval of $(l_0,\, +\infty)$, and $(\Delta)^{\alpha/2}u|_{(-l,l)}=f(x)$, with $f(x)$  continuous in $[-l,\, l]$. Looking for globally existing solutions, we shall need the respective Riesz potentials:
\begin{equation*}
u_{\alpha,f}^0(x)=\ds{\int_{l_0}^{\infty}}\frac{c_{1,\alpha}f(y)dy}{|x-y|^{1-\alpha}},\, u_{\alpha,f}(x)=\ds{\int_{-l}^{l}}\frac{c_{1,\alpha}f(y)dy}{|x-y|^{1-\alpha}},\,
\left( c_{1,\alpha}=\frac{\Gamma (\frac{1-\alpha}{2})}{2^{\alpha}{\sqrt{\pi}\Gamma (\frac{\alpha}{2})}}\right).
\end{equation*}
Clearly, for the existence of these potentials it is required that $0<\alpha<1$. Suggested from the possible singularities of the types ${\frac{1}{|x-l_0|^{1-\alpha}}}$ or 
$\frac{1}{|x\pm l|^{1-\alpha}}$, concerning respectively the first or the second equation above, we shall interesting in solutions $u\in S'=S'({\mathbb{R}^1})$ satisfying the relevant condition:
\begin{equation}
\label{A1}
a)\, (x-l_0)^{1-\alpha}u(x)\in L^{\infty}(l_0-1,l_0+1);\, b)\,(x\pm l)^{1-\alpha}u(x)\in L^{\infty}(-l-1,l+1).
\end{equation}
Below we shall use the notation $ |x-l^*|^{1-\alpha}u|_{x=l^*}$ for the limit (assumed existing)\\
 $ \ds{\lim_{x\to l^*}|x-l^*|^{1-\alpha}u(x)}$, $l^*\in\mathbb{R}^1$. Consider now the following boundary value problems of Dirichlet type:
\begin{equation}
\label{A2}
(-\Delta)^{\alpha/2}u|_{(l_0,+\infty)}=f(x);\, |x-l_0|^{1-\alpha}u|_{x=l_0}=c_0\, (c_0 =const\in\mathbb{R}^1).
\end{equation}
\begin{equation}
\label{A3}
(-\Delta)^{\alpha/2}u|_{(-l,l)}=f(x);\, |x\pm l|^{1-\alpha}u|_{x=\mp l}=c_{\mp}^0\, (c_{-}^0,\,  c_{+}^0=const\in\mathbb{R}^1).
\end{equation}

Next, the question for resolving the problems is discussed separately but in a common framework. The relevant two assertions give the essence of the needed answer.
\begin{prop}
\label{prA1}
For $f(x)$, continuous in $[l_0,+\infty )$  and vanishing out of $[x_1,x_2]\subset[l_0,+\infty )$, and an arbitrary constant  $c_0$ problem (\ref{A2}) has a unique solution $u\in S'$  satisfying condition (\ref{A1}.a), expressed by the formula:
\begin{equation}
\label{A4}
u(x)=\frac{c_0}{|x-l_0|^{1-\alpha}}+\int_{l_0}^{+\infty}\frac{c_{1,\alpha}f(y)dy}{|x-y|^{1-\alpha}},\, (x\in\mathbb{R}^1).
\end{equation}
\end{prop}

\begin{prop}
\label{prA2}
For  $f(x)$-- continuous function in $[-l,l]$  and  $c_{-}^0$, $  c_{+}^0$ --  arbitrary constants problem (\ref{A3}) has a unique solution $u\in S'$  satisfying conditions (\ref{A1}.b), which is present by the formula:
\begin{equation}
\label{A5}
u(x)=\frac{c^0_{-}}{|x+l|^{1-\alpha}}+\frac{c^0_{+}}{|x-l|^{1-\alpha}}+\int_{-l}^{+l}\frac{c_{1,\alpha}f(y)dy}{|x-y|^{1-\alpha}},\, (x\in\mathbb{R}^1).
\end{equation}
\end{prop}

Sketch of proofs:
Suppose  $u\in S'$ is a solution of the equation from (\ref{A2}), satisfying condition (\ref{A1}.a), i.e. $(-\Delta)^{\alpha/2}u=f^0[f]$ in $S'$,  and 
$(-\Delta)^{\alpha/2}[u-u_{\alpha,f}^0]=0$ on $\mathbb{R}^1\setminus\{l_0\}$, therefore $(-\Delta)^{\alpha/2}[u-u_{\alpha,f}^0]=C_0\delta(x-l_0)$, with a constant $C_0$, because of condition (\ref{A1}.a). More accurately, according to the known properties of the compactly supported distributions $u\in S'$ (\cite{Hermander}), instead of 
$C_0\delta(x-l_0)$  it should be taken a sum of the type $\ds{C_0\delta(x-l_0)+\sum_{m=1}^{N}C_m\delta^{(m)}(x-l_0)}$. 
However condition (\ref{A1}.a) yields  $C_m=0$ ( $m=1,2,\dots, N$). Next, as in the Introduction, by applying the Fourier transform to equation\\ 
$\ds{(-\Delta)^{\alpha/2}[u-u_{\alpha,f}^0]=C_0\delta(x-l_0)}$  we resolve it regarding $u$, finding the relation \\
$\ds{u=\frac{C_0 c_{1,\alpha}}{|x-l_0|^{1-\alpha}}+u_{\alpha,f}^0}$. Rewriting in details the potential  $u_{\alpha,f}^0$, we get the following general solution formula (with $C_0$  as a free constant):

\begin{equation}
\label{A6}
u(x)=C_{0}\frac{c_{1,\alpha}}{|x-l_0|^{1-\alpha}}+\int_{l_0}^{+\infty}\frac{c_{1,\alpha}f(y)dy}{|x-y|^{1-\alpha}},\, (x\in\mathbb{R}^1).
\end{equation}
In the case related to the final interval $(-l,l)$  we only have to use once more the above arguments and to solve now the equation 
$\ds{(-\Delta)^{\alpha/2}[u-u_{\alpha,f}^0]=C_{-l}\delta(x+l)+C_{l}\delta(x-l)}$.      
 The resulting expression for $u$ gives the next general formula:
\begin{equation}
\label{A7}
u(x)=C_{-l}\frac{c_{1,\alpha}}{|x+l|^{1-\alpha}}+C_{l}\frac{c_{1,\alpha}}{|x-l|^{1-\alpha}}+\int_{-l}^{+l}\frac{c_{1,\alpha}f(y)dy}{|x-y|^{1-\alpha}},\, (x\in\mathbb{R}^1).
\end{equation}

Substituting afterwards from (\ref{A6}), (\ref{A7}) respectively in the boundary conditions\\
 $\ds{|x-l_0|^{1-\alpha}u|_{x=l_0}=c_0}$, $\ds{|x\pm l|^{1-\alpha}u|_{x=\mp l}=c_{\mp}^0}$, we easily obtain the announced formulas (\ref{A4}), (\ref{A5}). Thus we actually get the uniqueness part (of Propositions \ref{prA1}, \ref{prA2}) and the existence one consists in the verification already known from the Introduction.   

 \section*{ Acknowledgments }
This work was partially supported by grant 80-10-53/10.05.2022 of the Sofia University Science Foundation.

\end{document}